\documentclass[utf8]{frontiersFPHY} 

\setcitestyle{square} 
\usepackage{url,hyperref,lineno,microtype,subcaption}
\usepackage[onehalfspacing]{setspace}
\usepackage[bottom]{footmisc}

\def\keyFont{\fontsize{8}{11}\helveticabold }
\def\firstAuthorLast{Zhang and Gr\"uneis} 
\def\Authors{Igor Ying Zhang\,$^{1}$, and Andreas Gr\"uneis\,$^{2,*}$}
\def\Address{
$^{1}$MOE Laboratory for Computational Physical Science, Shanghai Key Lab Mol Catalysis \& Innovative Materals, Collaborative Innovation Center of Chemistry for Energy Materials, Department of Chemistry, Fudan University, Shanghai 200433, China \\
$^{2}$Institute for Theoretical Physics, Vienna University of Technology, Wien, A-1040, Austria }
\def\corrAuthor{Andreas Gr\"uneis}
\def\corrEmail{andreas.grueneis@tuwien.ac.at}

\begin{document}
\onecolumn
\firstpage{1}

\title[CC theory in materials science]{Coupled cluster theory in materials science}

\author[\firstAuthorLast ]{\Authors} 
\address{} 
\correspondance{} 

\extraAuth{}

\maketitle

\begin{abstract}


The workhorse method of computational materials science is undeniably density functional theory (DFT) in
the Kohn-Sham framework of approximate exchange and correlation energy functionals. However, the need for
highly accurate electronic structure theory calculations in materials science
motivates the further development and exploration of alternative as well as complementary techniques.
Among these alternative approaches, quantum chemical wavefunction based theories and in particular coupled cluster theory hold the promise
to fill a gap in the toolbox of computational materials scientists. Coupled cluster (CC) theory provides a
compelling framework of approximate infinite-order perturbation theory in the form of an exponential of cluster operators
describing the true quantum many-body effects of the electronic wave function at a computational cost
that, despite being significantly more expensive than DFT, scales polynomially with system size.
The hierarchy of size-extensive approximate methods established
in the framework of CC theory achieves systematic improvability for many materials properties.
This is in contrast to currently available density functionals that often suffer from uncontrolled approximations that limit
the accuracy in the prediction of materials properties.

In this tutorial-style review we will introduce basic concepts of coupled cluster theory and
recent developments that increase its computational
efficiency for calculations of molecules, solids and materials in general. We will touch upon the
connection between coupled cluster theory and the random-phase approximation
that is widely-used in the field of solid state physics.
We will discuss various  approaches to improve the
computational performance without compromising accuracy. These approaches include large-scale
parallel design as well as techniques that reduce the prefactor of the
computational complexity. A central part of this article discusses the convergence of
calculated properties to the thermodynamic limit which is of significant importance for reliable
predictions of materials properties and constitutes an additional challenge compared to calculations
of large molecules.

We mention technical aspects of computer code implementations of periodic coupled cluster theories
in different numerical frameworks of the one-electron orbital basis; the
projector-augmented-wave formalism using a plane wave basis set and the numeric atom-centered-orbital (NAO)
with resolution-of-identity. We will discuss results and the possible scope of these implementations and
how they can help to advance the current state of the art in electronic structure theory calculations of materials.

\tiny
 \keyFont{ \section{Keywords:} quantum chemistry, computational materials science, coupled cluster, PAW, NAO} 
\end{abstract}

\section{Introduction}

The solution of the many-electron Schr\"odinger equation is at the heart of $ab-initio$ computational
materials science. Density functional theory (DFT) in the Kohn-Sham framework of approximate exchange
and correlation energy functionals is irrefutably the method of choice for the study of computational
materials science problems due to its good trade-off between accuracy and computational cost.
However, despite the great successes of DFT in the last decades, it remains difficult to improve
systematically upon the accuracy of currently available approximate density functionals, which sometimes fail even qualitatively
in cases where strong electronic correlation effects,
non-local van der Waals interactions or density-driven errors arising from the non-vanishing self-interaction
occur~\citep{Cohen2012}.
As a consequence, the community of $ab-initio$ computational materials scientists explores
the accuracy and computational efficiency of alternative methods to solve the many-electron problem,
such as Green's function based methods or many-electron wavefunction theories.
Quantum chemical wavefunction theories have the potential to treat electronic exchange and correlation
effects in a systematically improvable manner.
Their computational cost is in general significantly larger than that required for DFT calculations, which
limits their application to relatively small system sizes only. In this work we discuss recent efforts and
improvements that expand the scope of quantum chemical wavefunction theories and in particular that of
coupled cluster methods. Furthermore we outline implementation details and the relationship between the
coupled cluster ansatz and the random phase approximation briefly.
The latter method is becoming an increasingly efficient technique
to treat electronic exchange and correlation effects in solids and molecules.

Quantum chemical wavefunction theories include a wide range of methods that are capable of treating
weak as well as strong electronic correlation effects. In contrast to multi-reference methods,
single-reference coupled cluster theory is in general not suited to treat
strong correlation problems. However, coupled cluster theories have successfully been applied to
calculate a wide range of materials properties including (i) cohesive energies of
(molecular)
solids~\citep{Stoll2012,Schwerdtfeger2010-ir,Paulus1999,Yang640,Nolan2009,Booth2013,Gruber18,Gruneis11},
(ii) pressure-temperature phase diagrams~\citep{Gruber18b}
(iii) exfoilation energies of layered materials~\citep{Hummel2016,doi:10.1021/acs.jpclett.5b02174,Usvyat2018}
(iv) defect formation energies~\citep{grueneis2015explicitly} and
(v) adsorption and reaction energies of atoms and molecules on surfaces~\citep{Usvyat12,2018arXiv180806894T,
tsatsoulis2017,kubas2016,boese2016,voloshina2011}.
Furthermore equation of motion coupled cluster theories have been implemented to calculate excited
state and single-electron related properties including
electron-addition and removal energies in solids~\citep{McClain2017}.
As a model Hamiltonian for real metallic systems, the uniform electron
gas has also been studied extensively using CC and related
approaches~\citep{Freeman77,Bishop78,PhysRevB.93.235139,PhysRevLett.110.226401}.
This list of applications is by no means complete but illustrates the potential of coupled cluster methods
for computational materials science simulations.
The achieved accuracy depends on the employed level of truncation in the coupled cluster wavefunction ansatz.
Comparison to experimental findings and high-level quantum Monte Carlo (QMC) results for solids reveals that
coupled cluster singles and doubles theory including perturbative triples (CCSD(T)) achieves a similar level
of accuracy in solids as for molecular quantum chemistry applications, indicating that it is possible to obtain
results with chemical accuracy (1 kcal/mol) or even better for most energetic properties such as cohesive energies.
This level of accuracy is similar to the accuracy that can be achieved using QMC calculations although
a systematic assessment for solids as it was carried out for molecules is still missing~\cite{Nemec2010}.
However, in contrast to QMC, perturbative M\o ller-Plesset and
coupled cluster theories such as the perturbative triples approach are not suitable for the
treatment of metallic systems, and more work is needed to understand and possibly correct for these
deficiencies~\citep{PhysRevLett.110.226401}.
We stress that non-perturbative coupled cluster methods such as
coupled cluster singles and doubles (CCSD) will yield convergent
correlation energies even for metallic systems.
The applications referred to above have been obtained using a variety of different computer code implementations
of CC theories that can roughly be divided in three different categories:
(i) local schemes that are based on a localized occupied orbital manifold and 
take advantage of the short-rangedness of electronic correlation explicitly~\citep{Cryscor},
(ii) the incremental method that is extremely efficient for weakly interacting fragments,
such as molecular crystals~\citep{Stoll2012,Schwerdtfeger2010-ir,Paulus1999} and
(iii) canonical schemes that explicitly account for the translational symmetry of
periodic systems by employing delocalized Bloch orbitals~\citep{Gruber18,McClain2017}.
Recently another approach to coupled cluster theory using stochastic techniques has been developed and successfully
applied to a range of systems including the uniform electron gas~\cite{Thom2010,Neufeld2017,Spencer2018}.
In this article we will mostly focus on canonical schemes that employ delocalized Bloch orbitals and
we will discuss their advantages as well as disadvantages compared to other methods.

The general purpose of computationally expensive yet highly accurate electronic structure theories
in materials simulations is two-fold: (i) production of benchmark results and (ii) prediction of
materials properties without depending on uncontrolled approximations.
Benchmark results are very valuable because they can be used to further
test and improve upon the accuracy of computationally more efficient yet less accurate methods or adjust
parameters used in simplified models of real systems. Although experiment can also provide benchmark results, experimental
findings need to be corrected for the effect of lattice vibrations, finite temperature and possibly relativistic
contributions in order to be directly comparable to non-relativistic zero temperature electronic structure
theory calculations in the Born-Oppenheimer approximation.
As such it is often more efficient to employ theoretical benchmark results obtained using high-level theories.
Ideally it should be possible to perform benchmark calculations in a  black-box manner.
A black-box method allows for high-throughput calculations and enables non-expert users to
obtain reliable simulation results for real materials.
Compared to density functional theory,
many-electron wavefunction theory calculations are more difficult to apply in practice
because calculated properties such as the ground state energy converge
much slower with respect to the employed computational parameters such as $k$-point mesh density and basis set size,
often requiring extrapolation techniques that need to be checked carefully.
Recent progress makes it possible to accelerate the convergence with respect to such computational parameters
significantly and control approximations in an automated way such that coupled cluster methods become
more ``user-friendly''.
In this article we will review some of the most important and recent improvements that aim at expanding the scope
of coupled cluster theories to the field of computational materials science. Many methodological developments
are inspired by related techniques in the field of molecular quantum chemistry and
quantum Monte Carlo.

Due to the potential use of coupled cluster theory as an accurate benchmark tool in materials simulations,
it is also necessary to understand and revisit the influence of widely-used approximations in modern electronic
structure theory codes. Obviously, the precision of the underlying numerical approach to solve the Schr\"odinger
equation must not be lower than the accuracy of the employed many-electron theory.
Pseudopotentials, basis sets and even frozen core approximations that have become standard practice in DFT calculations
of solids can not necessarily be transferred directly to wavefunction theory calculations.
To this end it is important to assess the influence of such approximations by comparing wavefunction
theory calculations on different footings.
In this article we will report on two different implementations
of CC theory employing the numeric atom-centered orbital (NAO) and the projector
augment wave (PAW) framework.

The coupled cluster method accounts for many-electron correlation effects explicitly
using an exponential ansatz for the wavefunction.
The wavefunction amplitudes are obtained by solving a set of nonlinear equations that
can be derived using different methods including second quantization in combination with Wick's theorem,
Slater rules or diagrammatic techniques.
The latter are also popular in the field of Green's function based methods, where Feynman diagrams
serve as a representation of quantum field theoretical expressions of many-particle interactions.
Diagrammatic methods allow for characterizing the CC method as an approximate perturbation
theory that performs a summation of a certain type of diagrams to infinite order.
In a diagrammatic language the close relationship between coupled cluster theory and other approaches
such as the random-phase approximation (RPA) becomes more obvious for both ground and excited state properties
as discussed in detail in Refs.~\citep{Berkelbach2018,Scuseria2008}.
This relationship also implies that the RPA and CC theory share
computational characteristics such as a slow convergence with respect to the employed independent
particle basis sets and $k$-point mesh. Therefore methodological improvements such as basis
set extrapolation methods and finite size corrections can often be readily transferred between these
approaches.



\section{Theory and concepts}
\subsection{Coupled cluster theory}

Coupled cluster theory was initially proposed by Fritz Coester and Hermann
K\"ummel in the field of nuclear physics~\citep{Coester58,Coester60}. In
the 1960s Jiri Cizek and Josef Paldus introduced the method for electron correlation~\citep{Cizek66,Cizek71}
and since then it has become a widely-used electronic structure theory method for quantum chemical calculations
on systems that do not exhibit strong static correlation~\citep{Bartlett07}.
For a more detailed introduction to coupled cluster theory we refer the reader
to Refs.~\citep{Crawford07,Bartlett07,shavitt2009many,Helgie}.
Coupled cluster theory uses an exponential ansatz of cluster operators for the many-electron
wavefunction
\begin{equation}
|\Psi^{{\rm CC}} \rangle = e^{\hat{T}} |\Phi^{\rm HF}\rangle,
\end{equation}
where $\Phi^{\rm HF}$ is a single Slater determinant constructed from the Hartree-Fock (HF) one-electron orbitals that best
approximates the ground state energy of a many-electron system.
In passing we stress, however, that CC theory can in principle employ any single reference determinant although this can affect
the accuracy.
The cluster operator in the exponent is defined by $\hat{T} = \sum_{m=1}^{n} \hat{T}_m$, where 
$n$ corresponds to the order of the coupled cluster approximation.
$\hat{T}_m$ is an $m$-fold excitation operator that generates $m$-fold excited Slater
determinants $|\Phi_{i_1 .. i_m}^{a_1 .. a_m}\rangle$ multiplied by corresponding
amplitudes ($t_{i_1 .. i_m}^{a_1 .. a_m}$) when applied to the HF groundstate ($|\Phi^{\rm HF}\rangle$):
\begin{equation}
\hat{T}_m |\Phi^{\rm HF}\rangle= \sum_{\substack{i_1,..,i_m \in {\rm occ.} \\ a_1,..,a_m \in {\rm unocc.}}} t_{i_1 .. i_m}^{a_1 .. a_m} |\Phi_{i_1 .. i_m}^{a_1 .. a_m}\rangle.
\end{equation}
We will return to the discussion of the procedure for determining the amplitudes later.
The mean-field wavefunction obtained from HF theory serves as a single reference
determinant for the CC wavefunction and electronic correlation effects are accounted for explicitly
using excitation operators and corresponding amplitudes.
The indices $i$ and $a$ refer to occupied and unoccupied orbitals,
respectively. 
We stress that depending on the choice of orbitals and their symmetry, the wavefunction
amplitudes will reflect a different degree of sparsity. In the case of Bloch orbitals, as commonly used in periodic systems, the amplitudes
vanish unless the sum of the momenta of the occupied orbitals is equal to the sum of the momenta of the
unoccupied orbitals modulo a reciprocal lattice vector. We note that this property holds only for momentum-conserving Hamiltonians.

In coupled cluster singles and doubles theory the cluster operator is
approximated using $ \hat{T} \approx {\hat{T}_1+\hat{T}_2}$.
Due to the exponential ansatz of the CCSD wavefunction, the coefficients of
all $i$-fold excited Slater determinants with $i \ge 2$ are  approximated using antisymmetrized
products of single and double excitation amplitudes.
For two-electron systems, two-fold excited Slater determinants can be generated at most and therefore
CCSD theory becomes exact. Higher orders 
of coupled cluster theory become exact for systems with the corresponding number of electrons.
The cluster amplitudes $t_i^a$ and $t_{ij}^{ab}$ are obtained by inserting the wavefunction ansatz in the
Schr\"odinger equation, projecting on the left by $\langle \Phi_i^a | e^{-\hat{T}}$ and
$\langle \Phi_{ij}^{ab} | e^{-\hat{T}}$, and solving the amplitude equations
$\langle \Phi_i^a | e^{-\hat{T}} H e^{\hat{T}} | \Phi^{\rm HF} \rangle = 0$ and
$\langle \Phi_{ij}^{ab} | e^{-\hat{T}} H e^{\hat{T}} | \Phi^{\rm HF} \rangle = 0 $, respectively.
For practical computer implementations the latter equations are not useful but need to be recast
in expressions that depend explicitly on Hamiltonian matrix elements and amplitudes, which can be achieved
using further algebraic transformations including the Baker-Campbell-Hausdorff formula and Wick's theorem,
Slater rules or diagrammatic techniques~\citep{Crawford07,Bartlett07,shavitt2009many,Helgie}.

For most applications a good trade-off between computational cost and high accuracy is achieved
by truncating the cluster operator at doubles (CCSD) and accounting for the effect of triples in a perturbative manner.
This approach is referred to as CCSD(T) theory~\citep{RAGHAVACHARI1989479} and achieves chemical accuracy for the prediction of
reaction energies and barrier heights for a wide range of chemical reactions~\citep{Bartlett07,Helgie}.
We note that CCSD and CCSD(T) theory include all diagrams for the correlation energy
that occur in third- and fourth-order perturbation theory, respectively.
However, as a consequence of the single-reference approximation in CC theory, the treatment of strong correlation
problems is extremely limited.
Examples for strong correlation problems include molecular dissociation problems where the HF approximation to
the wavefunction fails dramatically due to the multideterminant nature of the true wavefunction.
Conventional single-reference CC theory are accurate only for wavefunctions that are dominated by single-reference determinants.
We note in passing that multireference coupled cluster theories aim at the treatment of strong correlation problems~\citep{Kohn2013,Evangelista2018}.

To establish a connection between coupled cluster methods and the random-phase approximation,
we now turn to coupled cluster doubles (CCD) theory. CCD theory approximates the cluster
operator using $ \hat{T} \approx {\hat{T}_2}$.
The cluster amplitudes $t_{ij}^{ab}$ are obtained by solving the quadratic amplitude equations
$\langle \Phi_{ij}^{ab} | e^{-\hat{T}_2} H e^{\hat{T}_2} | \Phi^{\rm HF} \rangle = 0 $ that in a canonical spin-orbital
basis read
\begin{equation}
\label{eq:t2amp}
\begin{aligned}
t_{ij}^{ab} = 
\frac{1}{\epsilon_i + \epsilon_j - \epsilon_a - \epsilon_b}
&  (
\langle ij || ab \rangle 
+ \langle cj || kb \rangle t_{ik}^{ac}
+ \langle ci || ka \rangle t_{jk}^{bc}
+ \langle cd || kl \rangle t_{lj}^{db} t_{ik}^{ac} \\
& + \frac{1}{2} \langle cd || ab \rangle t_{ij}^{cd} 
+ \frac{1}{2} \langle ij || kl \rangle t_{kl}^{ab} 
+ \frac{1}{4} \langle cd || kl \rangle t_{ij}^{cd} t_{kl}^{ab} \\
& - \langle cj || ka \rangle t_{ik}^{bc}
- \langle ci || kb \rangle t_{jk}^{ac}
- \langle cd || kl \rangle t_{lj}^{da} t_{ik}^{bc} \\
& + \frac{1}{2} \langle cd || kl \rangle  
\left [ t_{lj}^{ab} t_{ik}^{cd} - t_{li}^{ab} t_{jk}^{cd} + t_{ij}^{db} t_{kl}^{ac} -t_{ij}^{da} t_{kl}^{bc}  \right ]
 )
\end{aligned}
\end{equation}
In the above equation we use Einstein summation convention.
The indices $i,j,k$ and $l$ label occupied orbital indices, whereas
$a,b,c$ and $d$ label virtual orbital indices.
$\epsilon$ correspond to the HF one-electron energies and the anti-symmetrized electron repulsion
integrals (ERIs) are defined as
$\langle ij || ab \rangle  = \langle ij | r^{-1}_{12} | ab \rangle - \langle ij | r^{-1}_{12} | ba \rangle$
, where the ERIs are defined by
\begin{equation}
\langle ij | r^{-1}_{12} | ab \rangle = \int_{\Omega} \int_{\Omega} d {\bf x}_1 d {\bf x}_2 \frac{\chi^*_i ({\bf x}_1)
 \chi^*_j ({\bf x}_2) \chi_a ({\bf x}_1) \chi_b ({\bf x}_2) }{|{\bf r}_1-{\bf r}_2|}.
\end{equation}
In the above expression the  spin-orbitals $\chi$ depend on the space-spin coordinate
${\bf x}=({\bf r},\sigma)$ and the spatial coordinates are integrated over all space.
In section \ref{sec:implementation} we will return to the discussion of the evaluation of these integrals
using different frameworks of the independent particle basis sets.
Equation~(\ref{eq:t2amp}) is solved for the amplitudes in an iterative manner by updating the amplitudes
in every iteration using the right-hand side of Eq.~(\ref{eq:t2amp}). To accelerate convergence standard techniques
such as direct inversion of the iterative subspace (DIIS) can be employed~\citep{PULAY1980393}.
Once the amplitudes are obtained, the CCD correlation energy can be calculated by
\begin{equation}
E_c^{\rm CCD}= \frac{1}{4} \langle ij || ab \rangle t_{ij}^{ab}
\label{eq:ccdenergy}
\end{equation}

As a consequence of the large number of virtual orbitals, $N_v$, compared to the number of occupied orbitals, $N_o$,
needed to obtain converged correlation energies even when extrapolation techniques are employed, the computation and
storage of ERIs as well as their contraction with amplitudes constitute the main source of computational cost and
memory in canonical CC calculations. In terms of computational cost and memory the so-called particle-particle ladder term
$\langle cd || ab \rangle t_{ij}^{cd}$ exhibits the most unfavourable scaling with respect to the number
of virtual orbitals $\mathcal{O}(N_v^4 N_o^2)$.
In passing we note that a number of techniques have been suggested to reduce the computational cost of the
$\langle cd || ab \rangle t_{ij}^{cd}$ term in CC calculations~\citep{Hummel2017,Dutta2018} and
that local schemes can treat this term very efficiently by construction since it couples a single electron-pair to itself only.
The distribution of memory and computational load in parallel computer implementations as well as the
possible on-the-fly calculation of the required ERIs make it possible to study systems containing several hundreds
of virtual orbitals routinely. Furthermore modern tensor framework libraries greatly simplify the development of
compact and also fully parallel coupled cluster theory computer codes~\citep{solomonik2014}.

Equation~(\ref{eq:t2amp}) arranges the terms of the amplitude equations in the same manner
as Shepherd et al. in~\citep{Shepherd2014}, who investigated different
approximations to coupled cluster doubles theory for the uniform electron gas.
The various terms are labeled in agreement with the corresponding diagrams
such that the top line of Eq.~(\ref{eq:t2amp}) represents the driver and the ring terms, the second line
ladder terms, the third line crossed-ring terms, and the bottom line mosaics~\citep{Scuseria13}.
The quadratic amplitude equations of coupled cluster doubles theory couple all electron pairs of a
many-electron system using perturbation theory diagrams of a certain type to infinite order.
This is illustrated by the iterative solution for the amplitudes in Eq.~(\ref{eq:t2amp}):
in the first iteration it follows that
$t_{ij}^{ab} =\frac{\langle ij || ab \rangle }{ \epsilon_i + \epsilon_j - \epsilon_a - \epsilon_b} $.
In subsequent iterations the approximate first-order amplitudes are coupled via the corresponding
ladder, ring, crossed-ring, and mosaic terms to each other.
In this manner a summation of all possible couplings between different diagrams to infinite order is performed.
The included diagrams balance and account for important physical effects of many-electron systems.
All terms are accompanied by so-called exchange terms that
are needed to correct for exclusion principle violating (EPV) contributions and account
for the fermionic character of the many-electron wavefunction.
EPV contributions cause self-interaction errors that are not necessarily restricted to approximate DFT methods. 
Furthermore the magnitude of ring and ladder contributions to the electronic correlation energy is system dependent.
For homogeneous systems the dimensionality and electronic density plays a crucial role.
In the high-density limit the ring approximation becomes
exact for three-dimensional systems, whereas ladder terms become more important for
lower-densities and lower dimensions~\citep{Freeman77,Bishop78,Freeman78,Freeman83}.
The inclusion of ring diagrams is important to describe collective polarization effects and obtain
effectively screened inter-electronic interactions that yield convergent 
correlation energies in metals.

Coupled cluster theory is based on a systematically improvable
many-electron wavefunction ansatz and derived without using uncontrolled approximations.
However, various approximations to the amplitude equations
have been investigated and proven sometimes more accurate and even more stable than their parent method.
The distinguishable cluster approximation (DCSD)~\citep{Kats2013,Kats2015} demonstrates
that disregarding certain diagrams in the amplitude equations and reweighting others such that the resultant
method remains exact for two-electron systems yields results that clearly outperform CCSD in terms of
accuracy. Furthermore it is possible to show that by a change of representation DCSD exhibits a computational
complexity that scales with respect to system size
only as $\mathcal O(N^5)$ instead of $\mathcal O(N^6)$ as in the case of CCSD~\citep{Narbe2018}.
This illustrates that the key to efficient and accurate many-electron perturbation theories 
lies in carefully chosen and well balanced truncations of the many body perturbation series.
In this regard the random-phase approximation is a showcase for an efficient and reasonably accurate perturbation theory.


\subsection{The random-phase approximation and its connection to CC theory}

The random phase approximation (RPA) to the correlation energy dates back
to the 1950s. It was first introduced by Macke to predict convergent
correlation energies~\citep{macke_1950} in the uniform electron gas and was 
also developed by Bohm and Pines\citep{pines_collective_1952} for the
collective description electron interactions.
In the field of $ab-initio$ computational materials science the exact-exchange
plus correlation in the random-phase approximation has attracted
renewed and widespread interest in the last two decades~\citep{Paier2012}. 
This is due to the fact that computationally increasingly
efficient implementations have become available and that this method is capable of
describing all interatomic bonding situations reasonably well:
ionic, covalent, metallic, and even van der Waals bonding.
The computational complexity can even be lowered to $\mathcal O(N^3)$ in real space
formulations~\citep{kaltak_2014}.
Thus, the complexity of an RPA calculation does not exceed that of a
canonical hybrid density functional theory calculation, the prefactor is however considerably
larger.
The RPA correlation energy can be derived from many-electron Green’s function theory, 
or using the adiabatic-connection fluctuation-dissipation theorem (ACFDT),
or from coupled cluster theory.

As shown in~\citep{Scuseria2008}, it is possible to transform the RPA equations, that are usually expressed in a
general eigenvalue problem, to a quadratic Riccati equation that reads
\begin{equation}
\begin{aligned}
t_{ij}^{ab} = 
\frac{1}{\epsilon_i + \epsilon_j - \epsilon_a - \epsilon_b}
&  \left (
\langle ij | ab \rangle 
+ \langle cj | kb \rangle t_{ik}^{ac}
+ \langle ci | ka \rangle t_{jk}^{bc}
+ \langle cd | kl \rangle t_{lj}^{db} t_{ik}^{ac} 
 \right ).
\end{aligned}
\label{eq:rpat2amp}
\end{equation}
%
We stress that in the above equation $\epsilon$ correspond to the DFT one-electron energies.
Once the amplitudes are obtained, the RPA correlation energy can be calculated by
\begin{equation}
E_c^{\rm RPA}= \frac{1}{2} \langle ij | ab \rangle t_{ij}^{ab}
\label{eq:rpaenergy}
\end{equation}
Although the above formulation does not allow for an efficient computer implementation of the RPA, it illustrates that
the RPA and CCSD are closely related.
In the rings-only approximation, the second, third and fourth lines of Eq.~(\ref{eq:t2amp}) are disregarded.
Furthermore the random-phase approximation includes the direct rings only.
This implies that instead of using the (double bar) anti-symmetrized integrals,
only $\langle ij | ab \rangle $ integrals are employed in the RPA amplitude and energy equations,
making it necessary to employ a different prefactor in the correlation energy expression to stay consistent with
many-body perturbation theory.
Consequently, RPA can not be viewed as a wavefunction theory although it can be obtained from the CC amplitude equations
as explained above. The close relationship between these approaches has motivated a number of post-RPA corrections, of which
we will discuss only a small selection.



In the RPA, the magnitude of total correlation energies is
significantly overestimated and binding energies are systematically underestimated compared
to experiment even for weakly interacting systems.
The origin of these shortcomings can in some cases be understood by comparing the RPA amplitude and energy
expressions in Eqs.~(\ref{eq:rpat2amp}) and (\ref{eq:rpaenergy}) to the coupled cluster doubles amplitude
and energy expressions in Eqs.~(\ref{eq:t2amp}) and (\ref{eq:ccdenergy}), respectively.

The lack of exchange-like terms in the RPA leads to the inclusion of EPV contributions
(such as $\langle ii | ab \rangle t_{ii}^{ab}$) causing self-correlation errors; for example,
RPA yields non-zero electron correlation energies for one-electron systems. 
This has motivated the introduction of various post-RPA corrections such as second-order
screened exchange (SOSEX), AC-SOSEX or approximate exchange kernel methods (AXK).
The inclusion of SOSEX to the random-phase approximation dates back to Monkhorst and Freeman in the 70s
\citep{monkhorst_1973,Freeman77}.
Freeman showed in Ref.~\citep{Freeman77} that absolute correlation energies of the uniform electron gas get
significantly improved when compared to more accurate methods if SOSEX correlation energy contributions
are included.
Furthermore the underestimation of cohesive energies calculated in the RPA originates from the fact that
EPV contributions are larger for spin-polarized atoms than for non-spin-polarized solids.
Therefore the RPA+SOSEX approximation yields more accurate cohesive energies compared to experiment than the
RPA. We note in passing that similar findings have been obtained using related corrections to
the RPA such as AXK~\citep{Chen2018}.

Another important difference between RPA and coupled cluster theory is the choice of the reference determinant
and orbital energies. In the framework of ACFDT-RPA, the KS-DFT reference determinant is well justified.
However, from the perspective of CC theory the KS-DFT orbitals violate the Brillouin's theorem and the
correlation energy expression would therefore have to include terms that depend on the off-diagonal
Fock matrix elements.
These considerations have motivated the renormalized singles excitations (rSE) and related methods~\citep{Paier2012,ren2013,Klimes2015}.
The inclusion of these contributions improves upon the description of weakly-bound molecules
and solids compared to the RPA.

\subsection{Size extensivity and the thermodynamic limit}

In this section we discuss size extensivity and methods that accelerate the convergence
to the  thermodynamic limit (TDL).
Both concepts are highly relevant for the application of quantum chemical methods to solids.
In contrast to molecular systems, properties of solids or surfaces need to
be calculated in the TDL in order to allow for a direct comparison to experiment
and truly model an infinite periodic system.
The TDL can be approached in different ways using; for example,
(i) sampling of the Brillouin zone with increasingly dense $k$-point meshes and in periodic boundary conditions,
(ii) studying increasingly large supercells in periodic boundary conditions, or employing
(iii) increasingly large clusters with open boundary conditions and/or embedding methods.
Once the TDL is approached with respect to the number of $k$-points or the number of atoms in the cluster,
intensive properties such as the correlation energy per atom are converged to a
constant value.

An important advantage of truncated coupled cluster theories compared to truncated configuration interaction
methods is their size extensivity.
Size extensivity is a concept of particular importance in quantum chemistry, which judges
if the calculated quantities have the correct asymptotic size dependence or not.
For extensive quantities, like the (correlation) energy, a given size extensive method  should yield the
asymptotic $N^1$ dependence with $N$ is the number of unit cells or wave vector sampling points in the
Brillouin zone~\citep{hirata:2011A}.  Obviously, the methods with incorrect asymptotic N$^{\alpha}$
dependence of $\alpha<1$, like the truncated configuration interaction methods, lead to the total energy per
unit cell equal to that of the HF mean-field approximation, which is clearly useless
for condensed-matter systems. The size extensivity of coupled cluster theories can also be understood via 
either the diagrammatic criteria~\citep{bartlett:1981A} or the supermolecule criterion~\citep{szabo:1996A}. 
Moreover, it was argued that approximate post-HF correlation methods cannot capture the variational and 
size-extensive properties simultaneously~\citep{hirata:2014A}.

The TDL is approached as the number of particles becomes infinite in
the simulation (super-)cell while the density is kept constant.
Once the TDL is approached, correlation energies per atom need to be converged to a constant for periodic systems,
corresponding to $\alpha=1$.
Finite size errors are defined as the difference between the TDL and the
finite simulation cell results.
However, converging calculated properties with respect to the system size can be very slow,
requiring substantial computational resources due to the steep scaling of
the computational complexity of coupled cluster methods with respect to system size.
We stress that many properties such as the binding energy of molecules on  surfaces
converge even slower than their counterparts calculated on the level of mean-field theories such as DFT.
Correlated wavefunction based methods capture long-range electronic correlation effects; for example dispersion interactions,
explicitly. Although the respective  contribution to the electronic correlation energy can be small,
the accumulation of such interactions can become a non-negligible contribution to the
property of interest. Various different strategies have been developed to correct for finite size errors
that are defined as the difference between the TDL and the finite simulation cell results.
These strategies often involve extrapolation methods or range-separation techniques.
Local theories that employ correlation energy expressions depending on localized electron pairs,
can approximate correlation energy contributions of long-distant pairs using computationally more
efficient yet less accurate theories. Alternatively local theories can account for electron
pairs that are disregarded based on a distance criterion by using
an $R^{-6}$-type extrapolation~\citep{Usvyat12}.
Canonical implementations of periodic post-HF methods employ scaling laws for extrapolations to the TDL
that are based on an analogue rationale~\citep{Booth2013,Del_Ben2013-or,McClain2017}.
Auxiliary field quantum Monte Carlo theory employs finite size corrections that
are based on parametrized density functionals obtained from finite uniform electron gas simulation
cells~\citep{PhysRevLett.100.126404}.

\begin{figure}[h!]
\begin{center}
\includegraphics[width=8cm]{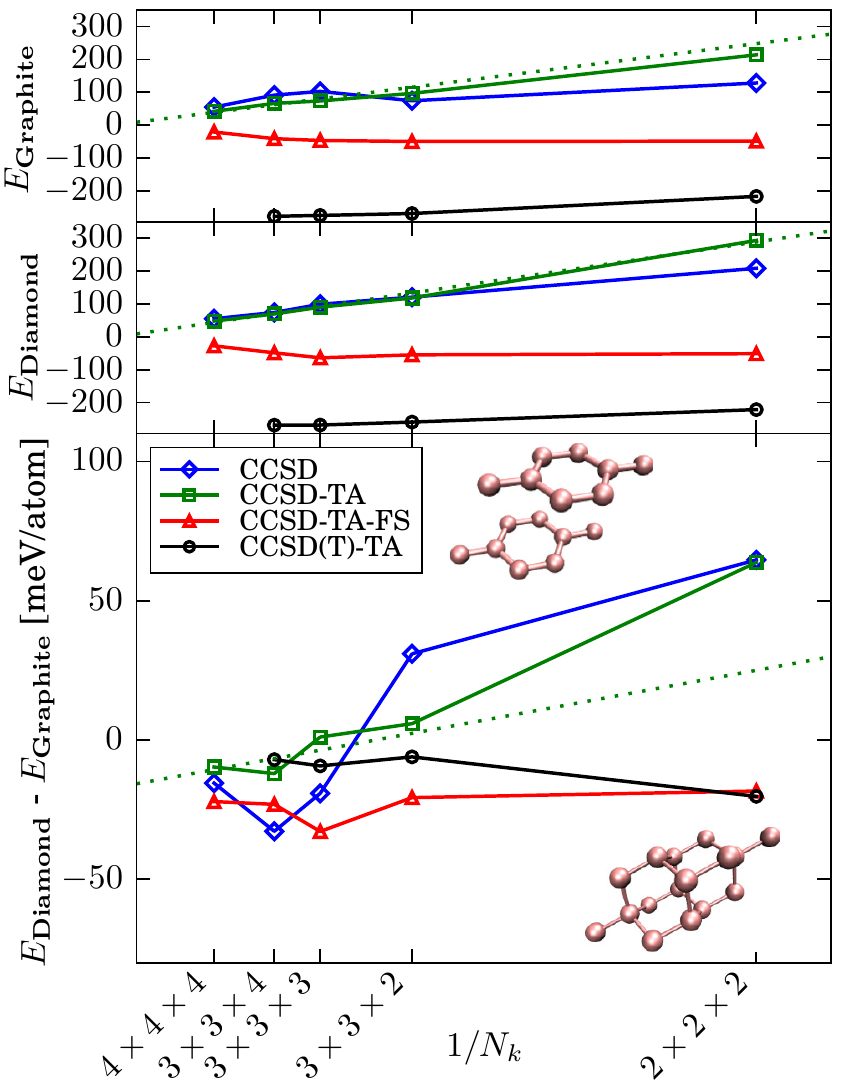}
\end{center}
\caption{ 
This figure has been taken from Ref.~\citep{Gruber18}.
Convergence of the correlation energy (difference) to the thermodynamic limit with respect
to the number of $k$-points ($N_k$) for
carbon graphite (top panel), carbon diamond (middle panel) and the difference between
carbon diamond and graphite (bottom panel). 
The dotted line represents the linear fit of the
uncorrected values. (T) corrections (black) are on
top of CCSD-TA-FS energy obtained using a 4$\times$4$\times$4 k-point
mesh (red). Zero-point energies are included, and they stabilize
graphite compared to diamond by 9~meV/atom.}\label{fig:tdl}
\end{figure}

We stress that the problem of slow thermodynamic limit convergence and concomitantly
large finite size errors is a common feature of Quantum Monte Carlo (QMC)
and many-electron quantum chemical methods due to their explicit treatment of electronic
correlation effects.
Recently structure factor interpolation and twist averaging (TA) techniques have been introduced in
Ref.~\citep{Gruber18} to correct efficiently for finite size errors in CC calculations of solids.
These techniques are inspired by related methods
used in QMC calculations~\citep{Chiesa2006,Holzmann2016}.
These finite size (FS) corrections allow for achieving thermodynamic limit results of solids and surfaces
using quantum chemical wavefunction theories in a very efficient manner~\citep{Gruber18,Liao2016},
reducing the computational cost significantly.
As shown in Figure~\ref{fig:tdl},
it is possible to achieve a much more rapid TDL convergence for simple solids such as carbon graphite
and diamond by virtue of these corrections. Furthermore the fact that the converged energies
per atom remain constant as the $k$-mesh density is increased,
reflects that CC theories are size extensive.

\section{Implementation and application}
\label{sec:implementation}

\subsection{Projector-augmented wave method}

In this section, we briefly review the calculation of two-electron repulsion integrals in the
framework of the projector augmented wave (PAW) method using a plane wave basis set. These integrals are
the most important quantities in addition to the cluster amplitudes for coupled cluster theory calculations.
In the PAW method, the all-electron orbitals ($| \psi_n \rangle$)
are obtained from the pseudo orbitals ($| \tilde{\psi}_n \rangle$)
using a linear transformation~\citep{Blochl1994},
\begin{equation}
| \psi_n \rangle = | \tilde{\psi}_n \rangle + 
                   \sum_i ( | \varphi_i \rangle - | \tilde{\varphi}_i \rangle ) 
                   \left \langle \tilde{p}_i | \tilde{\psi}_n \right \rangle.
\label{eq:paw}
\end{equation}
The index $n$, labelling the orbitals $\psi$,
is understood to be shorthand for the band index and the Bloch wave
vector $k_n$, while the index $i$ is a shorthand for the atomic site $R_i$,
the angular momentum quantum numbers $l_i$ and $m_i$, and an additional index $\epsilon_i$
denoting the linearization energy. The wave vector is conventionally chosen to lie
within the first Brillouin zone.
The pseudo orbitals are the variational quantities of the PAW method
and are expanded in reciprocal space using plane waves,
$\langle {\bf r} | \tilde{\Psi}_n \rangle = 
      \frac{1}{\sqrt{\Omega}} \sum_{{\bf G}} C^n_{{\bf G}}  e^{i({\bf k}_n+{\bf G}) {\bf r}}
\label{eq:pw}
$.
In this framework, it is possible to evaluate two-electron repulsion integrals approximately using the following expression:
\begin{align}
\langle ij | {r}_{12}^{-1} | ab \rangle=\sum_{\bf G} \rho_{i}^{a}({\bf G})  \tilde{{v}}({\bf G})  {\rho^\ast}_{b}^{j}({\bf G}),
\label{eq:erirec}
\end{align}
where $\tilde{v}$ is the diagonal Coulomb kernel in reciprocal space $\frac{4 \pi}{{\bf G}^2}$.
We note the the reciprocal Coulomb kernel exhibits a singularity a ${\bf G}=0$. We employ a correction for
the Coulomb kernel at the singularity that is obtained using a scheme introduced by Gygi and Baldereschi~\citep{Gygi1986}.
In the present PAW implementation, the Fourier transformed codensities $\rho_{i}^{a}({\bf G})$ are approximated
using Eq.~(2.87) of Ref. \citep{thesisHarl} as originally implemented by Kresse {\it et al.} for the
calculation of correlation energies within the random phase approximation~\citep{Harl2008}.
The codensities and the diagonal Coulomb kernel make it possible to
calculate the ERIs in a computationally efficient on-the-fly manner.
We note that the memory footprint of the transformed codensities scales cubic with respect to the system size.
Furthermore it is also possible to further reduce the dimension of the auxiliary plane-wave
index $\bf G$ significantly for system where the large number of degrees of freedom provided
by the plane-waves are not needed to describe the employed codensities. A significant down-folding
of the plane wave basis set size is possible without compromising accuracy for calculations of surfaces
as well as atoms or molecules in a box. A convenient and computational efficient method to achieve this downfolding
is based on singular value decomposition and outlined in Ref.~\citep{Hummel2017}.
We note that our more recent implementation of CC theory that calculuates the ERIs
according to Eq.~(\ref{eq:erirec}) (cc4s) employs the cyclops tensor framework
(CTF,~\citep{solomonik2014}) and will be released in the near future.

\subsection{Numeric atom-centered orbital framework}
The implementation of quantum-chemistry methods in the numeric atom-centered orbital 
(NAO) framework utilizes  the resolution-of-identity (RI) technique (also known as 
'density fitting') to calculate the two-electron 
repulsion integrals which scale as $N^4$ in memory with respect to system size $N$. 
The key idea is to decompose the four-rank repulsion integrals in terms of
three-rank tensors with an $N^3$ scaling of memory and is therefore better suited to be pre-stored:
\begin{equation}
\label{eq:RIM}
\langle rs | {r}_{12}^{-1} | pq \rangle \approx\sum_{\mu}m_{pr}^{\mu}m_{qs}^{\mu},
\end{equation}
where the index $\mu$ runs over an auxiliary basis $\{P_{\mu}(\boldsymbol{r})\}$, 
and $m_{pr}^{\mu}$ is a decomposed three-rank tensor in the molecular orbital basis. 
In RI-V approximation, $m_{pr}^{\mu}$ is determined by directly
minimizing the errors in the repulsion integral of atomic orbitals using Eqs.~(47 and 55) of 
Ref.~\citep{ren:2012A}. A hybird-RI algorithm has be designed to balance the computing cost
and communication in the CCSD(T) implementation for molecules in the Fritz Haber Institute 
\emph{ab initio} molecular simulation (FHI-aims) package. Together with a domain-based
distributed-memory strategy, it allows for an effective utilization of the
quickly increasing memory bandwidth of today's supercomputers to avoid the on-the-fly
disk storage and minimize interconnect communication, particularly for the tensor contraction in
the evaluation of the particle-particle terms. As a result, 
an excellent strong scaling can be achieved up to over 10,000 cores~\citep{igor:2018A}. 
The parallel efficiency is competitive with the CC implementations in state-of-the-art high-performance
computing computational chemistry packages.

Furthermore, a local variant of RI-V, namely RI-LVL, has been developed
in the numeric atom-centered orbital framework. Unlike the standard RI-V approximation in Eq.~\ref{eq:RIM}, 
RI-LVL expands the products of basis functions only in the subset of those auxiliary basis functions 
$\{\mathcal{P}(IJ)\}$ which are located at the same atoms of $I$ and $J$ as the basis functions of $\chi_{i}$ 
and $\chi_{j}$:
\begin{equation}
\label{eq:RILVL}
\langle rs | {r}_{12}^{-1} | pq \rangle \approx\sum_{\substack{\mu\lambda\in \mathcal{P}(RP)\\ \nu\sigma\in \mathcal{P}(SQ)}
}(rp|\lambda)L_{\lambda \mu}^{RP}(\mu|\nu)L_{\nu\sigma}^{SQ}(\sigma|sq),
\end{equation}
where the three-center $(rp|\lambda)$ and two-center $(\mu|\nu)$ integrals are defined as
\begin{equation}
(rp|\lambda) = \int_{\Omega} \int_{\Omega} d {\bf x}_1 d {\bf x}_2 \frac{\chi^*_r ({\bf x}_1)
 \chi_p ({\bf x}_1) \chi_\lambda ({\bf x}_2)}{|{\bf r}_1-{\bf r}_2|}, 
\end{equation}
and
\begin{equation}
(\mu|\nu) = \int_{\Omega} \int_{\Omega} d {\bf x}_1 d {\bf x}_2 \frac{\chi^*_\mu ({\bf x}_1)
 \chi_\nu ({\bf x}_2)}{|{\bf r}_1-{\bf r}_2|}, 
\end{equation}
respectively. $\boldsymbol{L}^{RP}=(\mu|\nu)^{-1}$ with $\mu$, $\nu\in\mathcal{P}(RP)$. 
This choice further reduces the memory scaling to $N^2$ without compromising the accuracy~\citep{igor:2015A}.
For molecules, it is a useful alternative to the standard RI-V,
which, however, is the only option for solids, since the memory consumption of RI-V quickly
becomes unaffordable with the increase of k-grid numbers in periodic boundary conditions~\citep{sergey:2015A}.
For MP2 and RPA, we demonstrated that the NAO-based periodic implementation can provide a smooth and consistent convergence towards the complete k-grid limit~\citep{igor:2018B}. The periodic CCSD in FHI-aims has been implemented,
but not yet fully optimized.

\section{Results}

\subsection{PAW-based coupled cluster applications}

\begin{figure}[h!]
\begin{center}
\includegraphics[width=6cm]{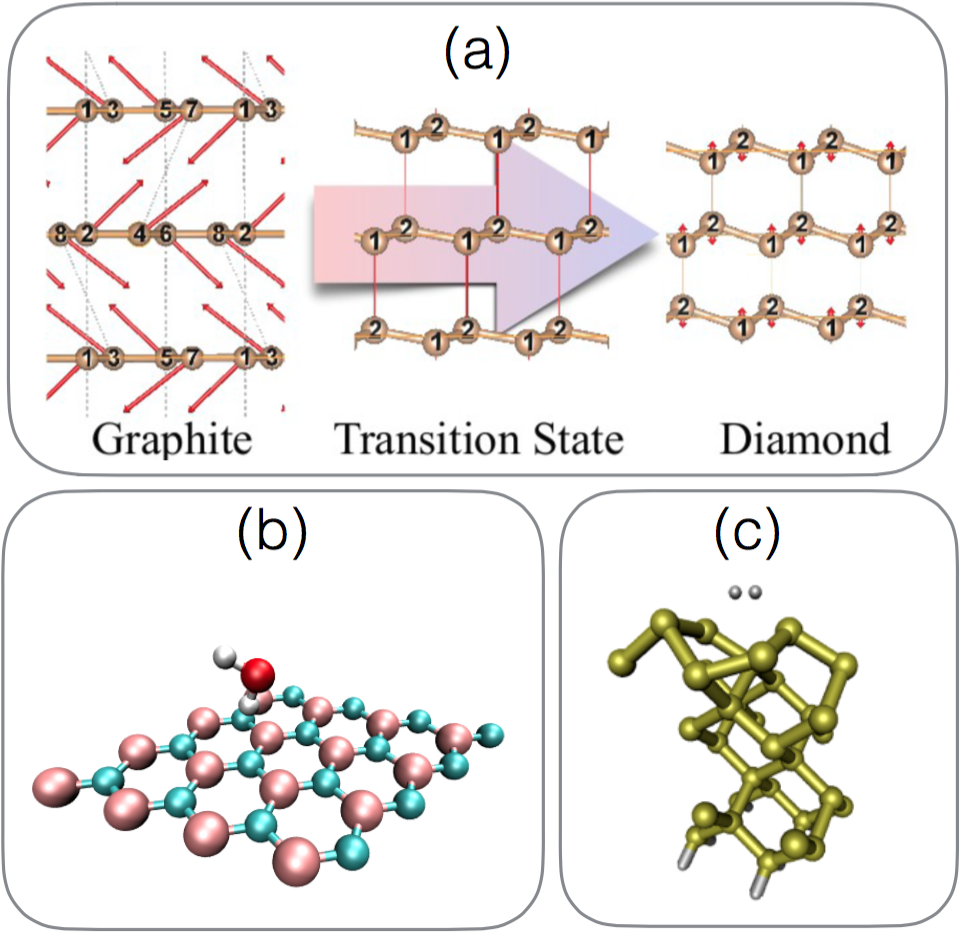}
\end{center}
\caption{
Overview of recent PAW-based periodic coupled cluster theory calculations:
\textbf{(a)} pressure-driven phase transition from carbon graphite to diamond\citep{Gruber18b},
\textbf{(b)} water adsorption on an $h-$BN sheet\citep{Gruber18} and
\textbf{(c)} dissociative hydrogen molecule adsorption on the Si(100) surface~\citep{2018arXiv180806894T}.
}\label{fig:paw_applications}
\end{figure}

We now turn to a brief overview of $ab-initio$ computational materials science studies that have
been performed using periodic coupled cluster theory.
This overview is by no means complete but reflects the general purpose and scope of the CC implementation 
based on the PAW formalism and its relation to other approaches such as the method of increments.
A selection of the investigated systems is depicted in Fig.~\ref{fig:paw_applications}.
PAW-based periodic CC calculations have initially been limited to insulating solids
with two-atomic unit cells such as the LiH crystal or the uniform electron gas simulation
cell~\citep{Gruneis11,Booth2013,PhysRevLett.110.226401}.
However, recent methodological improvements make the study of surfaces and supercells
containing approximately thirty atoms possible.
In particular the development of finite size corrections~\citep{Gruber18},
improved compact approximations to the
virtual orbital manifold~\citep{Gruneis11,booth2016}
and auxiliary basis sets~\citep{Hummel2017} have substantially expanded the scope
of periodic CC calculations. The advancement of explicitly correlated techniques in combination with a plane
wave basis set holds the promise to reduce the computational cost further in the near
future\citep{Gruneis2013-fk,Tenno2017}. 

\begin{table}[t]
  \begin{tabular}{ l | c  c  c  c  c   }
   & Periodic PAW$^a$ & Local Method$^b$ & Periodic GTOs$^c$ & Incremental Method$^d$  & Clusters$^e$  \\ \hline
    HF          &    3.583              &  3.591          &   3.592               & 3.591              &   3.589           \\
    MP2 corr.   &    1.187              &  1.125          &   1.200               & 1.131              &   1.182               \\
    CCSD corr.  &    1.326              &                 &                       &                    &   1.329              \\ 
    CCSD(T) corr.  & 1.383              &                 &                       & 1.334              &   1.381              \\ 
  \end{tabular}
  \caption{
  Calculated energy contributions to the cohesive energy of the LiH crystal as obtained using different theories and techniques.
  All energies in eV per LiH.
  $^a${Refs. \cite{Gruneis11,Booth2013} }.   
  $^b${Refs. \cite{Civalleri2010,Usvyat2011} }
  $^c${Ref. \cite{Del_Ben2013-or} }
  $^d${Ref. \cite{Stoll2012} }
  $^e${Ref. \cite{Nolan2009} }.
   }
  \label{tab:LiH}
\end{table}

\begin{table}[t]
  \begin{tabular}{ l | r r     }
                & Periodic PAW     & Local Method        \\ \hline
    HF          &   15             &  14                 \\ 
    MP2         &  233             & 238                 \\ 
    CCSD(T)     &  254             & 256                 \\ 
  \end{tabular}
  \caption{
  Calculated energy contributions to the adsorption energy of water on a (001) LiH surface model as obtained using different theories and techniques.
  The corresponding DMC estimate for the adsorption energy is 250 meV.
  All values are in meV and have been taken from Ref.\cite{tsatsoulis2017}.}
  \label{tab:waterLiH}
\end{table}

We first review studies that aim at assessing the precision of periodic CC calculations.
An important task of method development includes demonstrating
convergence of calculated properties such as the ground state energy with respect to the employed
computational parameters; for example, basis set and $k$-mesh, and achieve agreement with results obtained using
entirely different computational approaches.
In this regard the calculation of the cohesive energy of the LiH solid on the level of HF, MP2 and CC
theory has become common practice for benchmarking various implementations.
The first quantum chemical calculations of the LiH crystal were performed by Egil Hylleraas in 1930~\citep{hylleraas1930wellenmechanische}.
In 2009, Nolan $et~al.$ have presented MP2 and coupled cluster ground state energy calculations of the LiH
crystal converged with respect to basis set and system size through a combination of periodic and
finite-cluster electronic structure calculations~\citep{Nolan2009}.
These findings have served as a reliable reference for other approaches including (local) periodic
and incremental methods used to calculate MP2 or CC cohesive energies~\citep{Marsman2009-ah,Gruneis11,Stoll2012,Del_Ben2013-or,Usvyat2013-xi}.
Table~\ref{tab:LiH} summarizes the obtained cohesive energies, indicating the achieved level of precision
of the various calculations is better than chemical accuracy. Furthermore the CCSD(T) estimates agree with the experimental
value (4.98~eV/LiH) to within a few ten meV.
In a similar context, the adsorption energy of a single water molecule
on the LiH surface has been studied and results have been compared among different
theories and implementations to test their precision and accuracy~\citep{tsatsoulis2017}.
The corresponding adsorption energies are summarized in Table~\ref{tab:waterLiH} and demonstrate an agreement of a few meV between
different implementations.
Such benchmark calculations are useful despite their lack of reference to experiment.
However, to further assess the scope of an implementation it is necessary to investigate different
bonding situations.
Weakly bond molecular and rare-gas crystals have been studied intensively using the incremental method
with high accuracy~\citep{Schwerdtfeger2010-ir,Paulus1999}.
These systems are composed of many weakly interacting fragments,
making them ideally suited for an expansion of the correlation energy into few-body incremental contributions.
In contrast to the incremental method, a particular challenge for fully periodic
and canonical approaches originates from the long-rangedness of correlation energy contributions to the binding energy,
requiring dense $k$-mesh calculations or reliable finite size corrections~\citep{Gruber18}.
However, using recently proposed finite size corrections, it is possible to achieve well converged
CC cohesive energies of the neon solid in good agreement with results
obtained using the method of increments despite employing relatively coarse $k$-meshes~\citep{Gruber18,Schwerdtfeger2010-ir}.
The results obtained for the cohesive energy of the neon solid are summarized in Table~\ref{tab:Ne}.
The periodic CCSD(T) estimate for the cohesive energy agrees to within 3 meV/atom with experimental value of 27~meV/atom,
which has been corrected for zero-point fluctuations~\citep{Rosciszewski1999-me}.

\begin{table}[t]
  \begin{tabular}{ l | r  r     }
             & Periodic PAW (Ref.\cite{Gruber18}) & Incremental Method (Ref.\cite{Schwerdtfeger2010-ir})     \\ \hline
    MP2      &     17               &      19                          \\ 
    CCSD     &     19               &      22                         \\ 
    CCSD(T)  &     30               &      27                        \\ 
  \end{tabular}
  \caption{
  Calculated cohesive energy of the neon crystal as obtained using different theories and techniques.
  All energies in meV per atom. }
  \label{tab:Ne}
\end{table}

To benchmark the accuracy of periodic $ab-initio$ methods in a systematic manner it is common practice to
calculate cohesive energies, lattice constants and bulk moduli for a range of solids exhibiting different chemical bonding
including van-der Waals, ionic, covalent and metallic systems.
In molecular quantum chemistry similar calculations are routinely performed for a range of test sets
that reflect a range of different chemical bonding situations as discussed in the following section.
In solids, the systematic comparison to experimental findings, which have been corrected for beyond Born-Oppenheimer
approximation effects, allows to identify and understand systematic errors for different
levels of theory; for example, MP2 theory overestimates correlation energies for systems with small
band gaps, which results in an overestimation of corresponding cohesive energies~\citep{Gruneis2010a}.
However, the computational cost of coupled cluster calculations using the PAW-based periodic implementation
limits the number of benchmark systems so far.
The range of simple solids for which cohesive energies have been calculated in the complete basis set limit
on the level of CC theory include LiH (rock-salt), Ne (fcc), C (diamond), BN (zinc-blende) and
AlP (zinc-blende)~\citep{Booth2013,Gruber18}.
Furthermore phase transitions and energy differences of different
LiH, C and BN allotropes have also been investigated~\citep{Gruneis2015,Gruber18,Gruber18b}.
The study of pressure-temperature phase diagrams allows to investigate
the relative level of accuracy of electronic structure theories for different chemical bonds that are present
in the investigated phases. As such the predicted equilibrium phase boundaries often change significantly
with respect to the employed electronic structure theory.
In Ref.~\citep{Gruber18b} we have employed periodic CCSD(T) theory to calculate
relative enthalpies and pressure-temperature phase diagrams for C and BN allotropes.
In contrast to currently available DFT methods, quantum chemical wavefunction theories allow for achieving reliable
and systematically improvable benchmark results for the relative stability of C and BN allotropes.

\begin{table}[t]
  \begin{tabular}{ l | c  c   }
             & H$_2$O@graphene       & H$_2$O@$h-$BN      \\ \hline
    MP2      &                    &      -119                      \\ 
    CCSD     &                    &      -83 (-68)                      \\ 
    CCSD(T)  &     -87            &      -102(-87)                     \\ 
    DMC      &     -99$\pm$6      &      (-84)
  \end{tabular}
  \caption{
  Calculated adsorption energy of water on $h-$BN and graphene sheets. The numbers in parenthesis have been obtained without finite size
  corrections. All values have been taken from Refs.\cite{Hamdani2017,Gruber18,JGB2019}.}
  \label{tab:water}
\end{table}

Another promising area of application for quantum chemical wavefunction theories is the
study of surface science problems including molecular adsorption and reactions on surfaces.
CC theories are widely-used for predicting benchmark results for molecular gas-phase reaction energies
as well as activation barrier heights. A similar level of accuracy for molecular surface reactions can
currently only be achieved by QMC calculations and it would be advantageous to fully transfer quantum
chemical wavefunction based methods to this area. Currently available QC approaches are based 
mostly on embedding, fragment models, the incremental technique or finite-cluster
calculations~\citep{kubas2016,boese2016,voloshina2011,Usvyat12,libisch2012}.
Recently we have performed fully periodic calculations of adsorption energies for water on $h-$BN and
graphene sheets, confirming the expected level of accuracy~\citep{Gruber18,JGB2019}.
Table~\ref{tab:water} summarizes the results of the water adsorption energies and compares them to DMC findings that agree well.
Furthermore the dissociative H2 adsorption on the Si(001) surface has also
been investigated in Ref.~\citep{2018arXiv180806894T}.
The obtained results indicate that CC theory can predict accurate adsorption energies and study chemical reactions on surfaces in a reliable manner.
However, we stress that, despite the accurate findings discussed above, the level of accuracy always depends on the electronic structure
of the system and the level of many-electron wavefunction approximation.
Therefore high accuracy can only be achieved using CCSD(T) theory for systems that do not exhibit strong correlation effects.

\subsection{NAO-based coupled cluster applications}
As an alternative basis set choice instead of GTOs and plane waves, 
NAOs, in particular those with \emph{valence correlation consistency} namely NAO-VCC-nZ, 
hold the promise to provide improved description of advanced correlation methods
with the increase of the basis set size. NAO-VCC-nZ allows for extrapolating
the results of advanced quantum-chemistry methods to the complete-basis-set (CBS) 
limit. NAO-VCC-nZ was generated by minimizing frozen-core RPA total energies
of individual atoms from H to Ar. The consistent convergence of 
RPA and MP2 binding energies of small molecules is demonstrated by using NAO-VCC-nZ
basis sets in the original paper~\citep{igor:2013A}. 
The applicability of NAO-VCC-nZ for solids has been 
benchmarked comprehensively for MP2 and RPA calculations of cohesive energy, lattice constants,
and bulk modulus for representatives of first- and second-row elements and their
binaries with cubic crystal structures and various bonding characters~\citep{igor:2018B}. 
The generalization of the periodic RPA and MP2 implementation to CCSD(T) for solids is straightforward, 
but strong effort should be paid in order to achieve an efficient and practical implementation of the 
NAO-based periodic CCSD(T) method, which is still under development.

As the first step, the CCSD(T) has been implemented in FHI-aims for molecules, which has been used 
together with NAO-VCC-nZ basis sets to deliver accurate CCSD(T) results in the CBS limit for
manifold properties of great chemical or physical interest.
For 22 bio-orient weak interactions in the S22 test sets and 10 relative energies of 
cysteine conformers in the CYCONF test set, the CCSD(T)/CBS results
by using NAO-VCC-nZ repeat the up-to-date reference data based on GTO basis sets with the
deviation of less than 0.1 kcal/mol on average. 
For the first time, the high-level theoretical reference data at the CCSD(T)/CBS level 
has been generated for the 34 isomerization energies in the ISO34 test set in the NAO framework.
For the sake of benchmarking newly developed electronic-structure methods, the use of 
the CCSD(T)/CBS reference data allows for the comparison based on exactly the same
molecular geometry and immune to the experimental uncertainty~\citep{igor:2018A}.




\subsection{Outlook}

During the last years the scope of quantum chemical wavefunction theories and in
particular coupled cluster methods has been expanded significantly in the field of complex systems
and solids.
Fully periodic and canonical approaches have become increasingly efficient due to
methodological improvements that reduce the prefactor of the computational cost.
In this tutorial-style review we have given a brief overview of different frameworks for the
implementation of periodic CC theories and their applications.
Future work will focus on more systematic benchmark studies employing all recent methodological advancements
including finite size corrections and explicit correlation techniques for solids.
Furthermore it is expected that local approaches such as the use of truncated pair
natural orbitals~\citep{Nesse2009,Kubas16,Koeppl17} can also be transfered to solid state systems.
The remaining technical and theoretical challenges are, however, significant and include
the implementation of sparse tensor frameworks and the development of localization schemes
that work reliable for metals as well as for insulators and semiconductors. Furthermore
the calculation of additional properties such as gradients will eventually also be necessary
in order to allow for structural relaxation.
However, the high level of accuracy makes quantum chemical wavefunction theories and in particular
CC theory
a promising tool for future $ab-initio$ computational materials science studies that can complement
currently available DFT approaches and provide benchmark results.
As discussed in this article, possible areas of application include the study of pressure-driven
phase transitions, surface chemistry problems as well as optical properties of solids.

%

\section*{Conflict of Interest Statement}

The authors declare that the research was conducted in the absence of any commercial or financial relationships that could be construed as a potential conflict of interest.

%

\section*{Acknowledgments}
A.G. gratefully acknowledges support and funding from the European
Research Council (ERC) under the European Unions
Horizon 2020 research and innovation program (Grant Agreement No 715594).
Work at Shanghai was supported by the 14th Recruitment Program of Young Professionals for I.Y.Z.



\bibliographystyle{frontiersinHLTH_FPHY} 
\bibliography{grueneis} 

\end{document}